\documentclass[runningheads]{llncs}
\usepackage[T1]{fontenc}
\usepackage{graphicx}
\usepackage{amsmath}
\usepackage{subcaption}
\usepackage{multirow}
\usepackage{array}
\begin{document}

\title{Development of a Particle Bhatnagar-Gross-Krook Method for Rarefied Monatomic Gas Mixtures}
\titlerunning{Particle BGK Method for Rarefied Monatomic Gas Mixtures}

\author{
Inchan Kim\inst{1}\orcidID{0009-0008-6731-4924} \and
Joonbeom Kim\inst{1}\orcidID{0009-0001-3154-0152} \and
Woonghwi Park\inst{1}\orcidID{0009-0000-3969-372X} \and
Eunji Jun\inst{1}\orcidID{0000-0001-7403-6784}
}


\authorrunning{I. Kim et al.}

\institute{Korea Advanced Institute of Science and Technology, Daejeon 34141, Republic of Korea}

\maketitle

\begin{abstract}

The Direct Simulation Monte Carlo (DSMC) method has long served as a reliable numerical approach for simulating rarefied and non-equilibrium gas flows. Its computational cost, however, increases rapidly as the flow approaches the near-continuum regime. This limitation has motivated the development of kinetic model equations that replace binary collisions with simplified relaxation processes. This paper presents a particle multi-relaxation unified BGK (UBGK) model for monatomic gas mixtures. The model retains pairwise species relaxation and determines the target velocity, temperature, shear stress, and heat flux by matching the corresponding production terms to those of the Boltzmann equation. A stochastic particle algorithm is developed. The model is implemented in SPARTA, and two validation problems are considered: homogeneous relaxation and a Mach 3 flow over a cylinder. In both cases, the UBGK solution reproduces the DSMC reference behavior, including species-dependent relaxation and non-equilibrium shock structure. 


\end{abstract}

\section{Introduction}


Rarefied and non-equilibrium gas flows arise in high-altitude flight, atmospheric re-entry, electric propulsion systems, and microscale devices. DSMC is known to provide a stochastic solution of the Boltzmann equation using computational particles, and remains a standard method \cite{Bird1994-DSMC}. As the flow approaches the near-continuum regime, however, resolving molecular collision scales makes DSMC expensive. To overcome this limitation, kinetic model equations such as the Bhatnagar-Gross-Krook (BGK) and the Fokker-Planck (FP) equations provide an alternative by replacing binary collisions with simplified processes \cite{Jun2019-FPSlitOrifice,Jun2018-FPDSMC,Kim2025-FPInterpolation,Kim2024-FPInterpolation,Kim2024-FPDSMCHybrid,Kim2022-VHSFP,Kim2024-SOFP,Park2024-BGKCompare,Pfeiffer2018-BGKComparison,yao2023-SBGK+}.


The BGK equation models intermolecular collisions as a relaxation process toward a prescribed target distribution \cite{Bhatnagar1954-BGK}. In the original BGK model, the target distribution is Maxwellian. This simple formulation, however, yields a Prandtl number of unity. Several extensions have therefore been proposed to recover the correct transport behavior. The ellipsoidal statistical BGK (ESBGK) model modifies the shear stress relaxation, whereas the Shakhov BGK (SBGK) model modifies the heat flux relaxation. By combining these two approaches, the unified BGK (UBGK) model introduces auxiliary parameters that control both shear stress and heat flux relaxation \cite{Chen2015-UBGK}.


The extension of BGK models to gas mixtures requires an accurate description of momentum and energy exchange between different species, together with the corresponding mixture transport behavior. Existing mixture BGK models can be classified into single-relaxation and multi-relaxation formulations. Single-relaxation models collapse self- and inter-species interactions into a single effective relaxation operator, resulting in a relatively compact formulation \cite{Andries2002-SRMixBGK}. However, this simplified structure limits the explicit description of individual species-pair interactions. In contrast, multi-relaxation models retain separate relaxation terms for individual species pairs and therefore more closely resemble the pairwise structure of the mixture Boltzmann collision operator \cite{Bobylev2018-MRMixBGK}. This pairwise formulation requires appropriate relaxation rates and derivation of target properties to correctly reproduce inter-species momentum and energy exchange as well as mixture transport.


In the present study, a particle multi-relaxation UBGK model is presented for rarefied monatomic gas mixtures \cite{Kim2026-UBGKMixture}. The model allows pairwise relaxation between individual species and determines the corresponding target velocity, temperature, shear stress, and heat flux by matching the production terms with those of the Boltzmann equation. The model is assessed through homogeneous relaxation and a Mach 3 flow over a cylinder. These cases examine the model's capability to reproduce inter-species temporal relaxation and non-equilibrium flow behavior.

\section{Multi-relaxation UBGK Model for Gas Mixtures}


For an $N_S$-species mixture, the evolution of distribution function of species $\alpha$ follows the Boltzmann equation \cite{Bird1994-DSMC}:
\begin{equation}\label{eqn:boltzmann}
\begin{aligned}
    \frac{\partial{f^\alpha}}{\partial{t}} + \boldsymbol{c}^\alpha \frac{\partial{f^\alpha}}{\partial{\boldsymbol{x}}} = \sum_{\beta=1}^{N_S} S_{coll}^{\alpha\beta},
\end{aligned}
\end{equation}
The collision term $S_{coll}^{\alpha\beta}$ can be approximated with multi-relaxation BGK formulation, preserving species-pair structure as:
\begin{equation}\label{eqn:pairbgk}
\begin{aligned}
    S_{\mathrm{BGK}}^{\alpha\beta} = \nu_{\mathrm{BGK}}^{\alpha\beta} \left( f_{BGK}^{\alpha\beta} - f^\alpha \right),
\end{aligned}
\end{equation}
where $\nu_{\mathrm{BGK}}^{\alpha\beta}$ and $f_{BGK}^{\alpha\beta}$ are the pairwise relaxation rate and target distribution.


By extending the single-species UBGK model to a multi-relaxation formulation, the target distribution becomes  \cite{Kim2026-UBGKMixture}:
\begin{equation}\label{eqn:UBGK-TargetDistribution}
\begin{aligned}
    f_U^{\alpha\beta} = G'^{\alpha\beta} + S'^{\alpha\beta},
\end{aligned}
\end{equation}
\begin{subequations}
\begin{equation}\label{eqn:UBGK-TargetDistribution_ES1}
\begin{aligned}
    G'^{\alpha\beta} = \frac{n^\alpha}{\left( \mathrm{det} \left| 2 \pi \lambda'^{\alpha\beta} \right| \right)^{1/2}} \mathrm{exp} \left( -\frac{1}{2} C_i^{\alpha\beta} \left(\lambda'^{\alpha\beta} \right)^{-1} C_j^{\alpha\beta} \right),
\end{aligned}
\end{equation}
\begin{equation}\label{eqn:UBGK-TargetDistribution_ES2}
\begin{aligned}
    \lambda'^{\alpha\beta} = R^\alpha T^{\alpha\beta} \delta_{ij} + C_{ES} \frac{\sigma_{ij}^{\alpha\beta}}{\rho^\alpha},
\end{aligned}
\end{equation}
\end{subequations}
\begin{equation}\label{eqn:UBGK-TargetDistribution_S}
\begin{aligned}
    S'^{\alpha\beta} = G'^{\alpha\beta} \Bigg[ 
        C_S \frac{2 C_j^{\alpha\beta} q_j^{\alpha\beta}}{5 \rho^\alpha \left( R^\alpha T^{\alpha\beta} \right)^2 } \left( \frac{ C_j^{\alpha\beta} C_j^{\alpha\beta} }{ 2 R^\alpha T^{\alpha\beta} } - \frac{5}{2} \right)
    \Bigg],
\end{aligned}
\end{equation}
where $G'^{\alpha\beta}$ is the ESBGK distribution, $S'^{\alpha\beta}$ is the SBGK distribution, and $\lambda'^{\alpha\beta}$ represents the anisotropic stress tensor. The quantity $C_i^{\alpha\beta} = c_i^\alpha-u_i^{\alpha\beta}$ defines the thermal velocity relative to the relaxation velocity. The coefficients $C_{ES}$ and $C_S$ are the auxiliary parameters that control the shear stress and heat flux corrections. The parameters, namely $u_i^{\alpha\beta}, \allowbreak T^{\alpha\beta}, \allowbreak \sigma_{ij}^{\alpha\beta}$, and $q_i^{\alpha\beta}$, are the relaxation properties that constitute the UBGK model's target distribution function.


Closure of the mixture UBGK model is imposed through pairwise production-term matching,
\begin{equation}\label{eqn:matching}
\begin{aligned}
    P_{\mathrm{UBGK}}^{\alpha\beta}(\psi^\alpha) = P_{G13}^{\alpha\beta}(\psi^\alpha),
    \qquad
    \psi^\alpha \in \{ u_i^\alpha, T^\alpha, \sigma_{ij}^\alpha, q_i^\alpha \}.
\end{aligned}
\end{equation}
By reproducing these production terms, the model preserves the corresponding inter-species relaxation processes and recovers NSF-level transport behavior in the near-continuum regime \cite{Kim2025-DiMixFP}. The reference terms $P_{G13}^{\alpha\beta}$ are obtained from the mixture Boltzmann operator under the VHS molecule assumption \cite{Kim2025-FPMix,Kim2026-UBGKMixture}. Thus, $u_i^{\alpha\beta}$ and $T^{\alpha\beta}$ are determined to reproduce pairwise momentum and energy exchange, while $\sigma_{ij}^{\alpha\beta}$ and $q_i^{\alpha\beta}$ are determined to reproduce shear stress and heat flux production. The closed expressions for the four target properties are reported in Ref. \cite{Kim2026-UBGKMixture}.


A pairwise relaxation rate is introduced by modifying the G13 collision rate with a scaling factor yielding \cite{Kim2026-UBGKMixture}:
\begin{equation}\label{eqn:UBGK-relaxationrate}
\begin{aligned}
    \nu_{UBGK}^{\alpha\beta} = \nu_{G13}^{\alpha\beta} \sqrt{ \frac{2 m_r^{\alpha\beta}}{m^\alpha} },
\end{aligned}
\end{equation}
where $m_r^{\alpha\beta}$ is the reduced mass. With this modification, the UBGK relaxation rate becomes species-dependent, proportional to $\sqrt{1 / m^\alpha}$. Therefore, the relaxation rate increases as the molecular mass decreases, consistent with the expected physical behavior of gas mixtures \cite{Li2024-MRMixBGK}.


\section{Numerical Implementation}


Similar to the DSMC method, particle advection and relaxation steps are decoupled for the particle-based UBGK method \cite{Kim2026-UBGKMixture}. After particle advection, species macroscopic properties and all pairwise target properties are evaluated in each cell. Then, a particle of species $\alpha$ is selected for relaxation with a probability
\begin{equation}\label{eqn:prob}
\begin{aligned}
    P_{\mathrm{relax}}^\alpha = 1 - \exp \left( - \sum_{\beta = 1}^{N_S} \nu_{\mathrm{UBGK}}^{\alpha\beta} \mathrm{\Delta} t \right),
\end{aligned}
\end{equation}
and, conditional on relaxation, partner species $\gamma$ is chosen according to
\begin{equation}\label{eq:partner}
\begin{aligned}
    P(\gamma \, | \, \alpha) = \frac{\nu_{\mathrm{UBGK}}^{\alpha\gamma}}{\sum_{\beta = 1}^{N_S} \nu_{\mathrm{UBGK}}^{\alpha\beta}}.
\end{aligned}
\end{equation}
After a valid normalized velocity sample $\boldsymbol{\xi}^U$ is drawn from the UBGK distribution, the post-relaxation particle velocity is obtained by multiplying the relaxation thermal velocity and shifting by the relaxation velocity as:
\begin{equation}\label{eqn:UBGK-postrelaxation-velocity}
\begin{aligned}
    c_i^\alpha = u_i^{\alpha\gamma} + \xi_i^U \cdot v_{th}^{\alpha\gamma}.
\end{aligned}
\end{equation}

\section{Results and Discussion}


The mixture UBGK model is implemented in SPARTA, an open-source DSMC code developed by Sandia National Laboratories \cite{Plimpton2019-SPARTA}. Identical spatial and temporal discretizations are used for the mixture UBGK model and DSMC. The VHS parameters are listed in Table \ref{table:VHSparameter}.

\begin{table}[h]
\caption{VHS molecular parameters.}\label{table:VHSparameter}
\centering
    \begin{tabular}{>{\raggedright\arraybackslash}p{1.8cm}>{\centering\arraybackslash}p{2.4cm}>{\centering\arraybackslash}p{1.6cm}>{\centering\arraybackslash}p{1.6cm}>{\centering\arraybackslash}p{2.4cm}}
        \hline
        Species $(\alpha)$ & $d_{ref}^\alpha \,[\textrm{m}]$ & $\omega_{ref}^\alpha \,[\textrm{\,-\,}]$ & $T_{ref}^\alpha \,[\textrm{K}]$ & $m^\alpha \, [\mathrm{kg}]$ \\
        \hline
        \hline
        He & $2.331 \times 10^{-10}$ & $0.66$ & $273.15$ & $6.646 \times 10^{-27}$ \\
        Ne & $2.766 \times 10^{-10}$ & $0.66$ & $273.15$ & $3.351 \times 10^{-26}$ \\
        Ar & $4.170 \times 10^{-10}$ & $0.81$ & $273.15$ & $6.634 \times 10^{-26}$ \\
        Kr & $4.762 \times 10^{-10}$ & $0.80$ & $273.15$ & $1.392 \times 10^{-25}$ \\
        \hline
    \end{tabular}
\end{table}

\subsection{Homogeneous relaxation}


A homogeneous relaxation problem is investigated. With no spatial gradients or particle–boundary interactions present, the system evolves solely through intermolecular collisions. The working gas is a ternary $(\mathrm{Ne-Ar-Kr})$ mixture with equal number densities, $n^{Ne} = n^{Ar} = n^{Kr} = 5 \times 10^{19} \, \textrm{m}^{-3}$. The non-equilibrium initial conditions are listed in Table \ref{table:HomRel_IC}. The simulation employs a single computational cell of size $\mathrm{\Delta} x = 0.01 \, \mathrm{m}$, with a total of $1.5 \times 10^{8}$ simulated particles. The time step is set to $\mathrm{\Delta} t = 0.01 \, \tau_{mix} = 2.0 \times 10^{-7} \, \mathrm{s}$.

\begin{table}[!t]
\caption{\label{table:HomRel_IC}Initial conditions for homogeneous relaxation problem.}
\centering
    \begin{tabular}{>{\raggedright\arraybackslash}p{1.8cm}>{\centering\arraybackslash}p{1.8cm}>{\centering\arraybackslash}p{1.8cm}>{\centering\arraybackslash}p{3cm}>{\centering\arraybackslash}p{3.3cm}}
        \hline
        Species $(\alpha)$ & $u_x^\alpha(0)\,[\mathrm{m/s}]$ & $T^\alpha(0)\,[\mathrm{K}]$ & $\sigma_{ij}^\alpha(0)$ & $q_i^\alpha(0)$ \\
        \hline
        \hline
        Ne & $100$ & $200$ & $0.1\,\rho^{Ne} R^{Ne} T^{Ne}$ &  $0.3\,\rho^{Ne} \left( R^{Ne} T^{Ne} \right)^{1.5}$ \\
        Ar & $60$  & $230$ & $0.2\,\rho^{Ar} R^{Ar} T^{Ar}$ & $-0.1\,\rho^{Ar} \left( R^{Ar} T^{Ar} \right)^{1.5}$ \\
        Kr & $0$   & $310$ & $0.3\,\rho^{Kr} R^{Kr} T^{Kr}$ & $-0.2\,\rho^{Kr} \left( R^{Kr} T^{Kr} \right)^{1.5}$ \\
        \hline
    \end{tabular}
\end{table}

\begin{figure*}[t!]
	\centering
	\begin{subfigure}{0.485\textwidth}
		\centering
        \includegraphics[width=1.\linewidth]{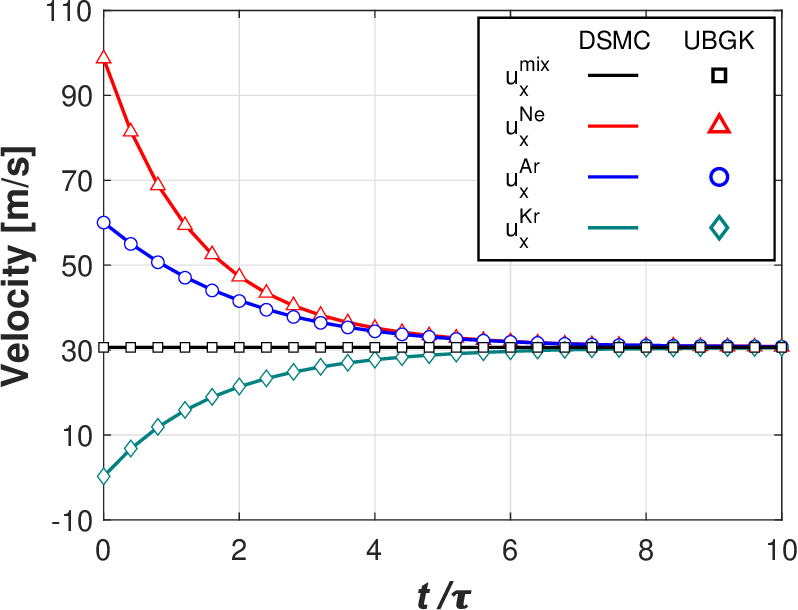}
            \caption{\label{subfig:1_Relax-Velo}Velocity.}
	\end{subfigure}
	\begin{subfigure}{0.485\textwidth}
		\centering
        \includegraphics[width=1.\linewidth]{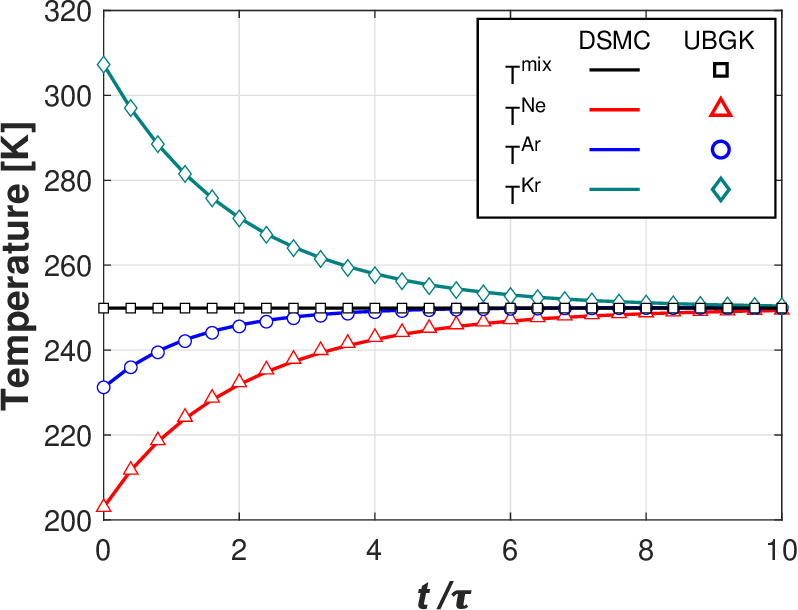}
            \caption{\label{subfig:1_Relax-Temp}Temperature.}
	\end{subfigure}
	\centering
	\begin{subfigure}{0.495\textwidth}
		\centering
        \includegraphics[width=1.\linewidth]{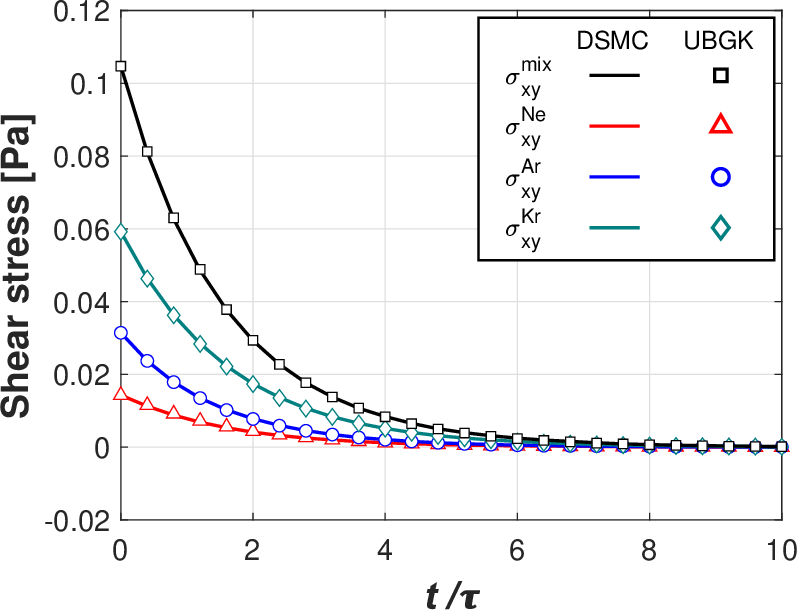}
            \caption{\label{subfig:1_Relax-Shear}Shear stress.}
	\end{subfigure}
	\begin{subfigure}{0.485\textwidth}
		\centering
        \includegraphics[width=1.\linewidth]{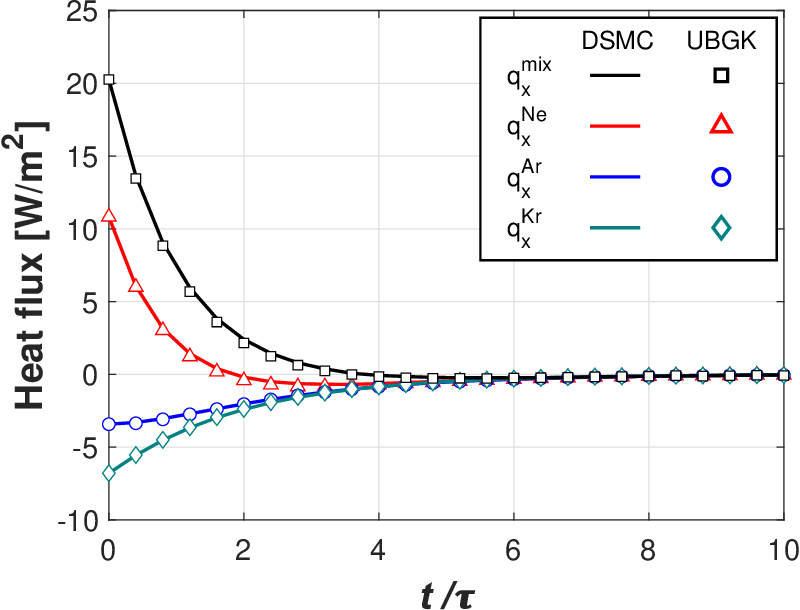}
            \caption{\label{subfig:1_Relax-heat}Heat flux.}
	\end{subfigure}
	\caption{\label{fig:1_Relax}Profiles of velocity, temperature, shear stress, and heat flux for the mixture, Neon, Argon, and Krypton.}
\end{figure*}


Figure \ref{fig:1_Relax} compares the UBGK and DSMC relaxations of $u_x^\alpha$, $T^\alpha$, $\sigma_{xy}^\alpha$, and $q_x^\alpha$ of the mixture as well as individual species. As shown in Figs. \ref{subfig:1_Relax-Velo} and \ref{subfig:1_Relax-Temp}, species velocities and temperatures relax toward the corresponding mixture values, since the system's total momentum and energy are conserved. In Figs. \ref{subfig:1_Relax-Shear} and \ref{subfig:1_Relax-heat}, $\sigma_{xy}^\alpha$ and $q_x^\alpha$ decay toward zero due to dissipation, as the system relaxes toward thermodynamic equilibrium. The UBGK results closely follow DSMC throughout the relaxation process without phase lag.

\subsection{Mach 3 flow over a cylinder}


A 2D cylinder flow is considered to assess the UBGK model under strong shock-induced non-equilibrium. The working gas is a ternary $(\mathrm{He-Ar-Kr})$ mixture with mole fractions $\left( \chi^{\mathrm{He}}, \chi^{\mathrm{Ar}}, \chi^{\mathrm{Kr}} \right) = \left( 0.2, 0.3, 0.5 \right)$. The cylinder diameter and wall temperature are $0.3048 \, \mathrm{m}$ and $300 \, \mathrm{K}$, respectively, and is assumed to be fully diffusive. The freestream conditions are $U_\infty = 675.37 \, \mathrm{m/s}$ and $T_\infty = 200 \, \mathrm{K}$, corresponding to $\mathrm{Ma} = 3$ and $\mathrm{Kn} = 0.02$. A $1 \, \mathrm{m} \times 0.8 \, \mathrm{m}$ half-body computational domain is discretized using a uniform $1000 \times 800$ grid with $\mathrm{\Delta} x = \mathrm{\Delta} y = \lambda_{mix} / 6$ and the adopted time step is $\mathrm{\Delta} t = 0.01 \, \tau_{mix} = 2 \times 10^{-7} \, \mathrm{s}$. The simulation is initialized with $200$ particles per cell, and the steady-state particle count is approximately $2.08 \times 10^8$. The flow conditions are summarized in Table \ref{table:2_Cyl-Condition}.

\begin{table}[t]
\caption{\label{table:2_Cyl-Condition}Flow conditions for a Mach 3 flow around a cylinder.}
\centering
\begin{tabular}{>{\centering\arraybackslash}p{1.4cm}>{\centering\arraybackslash}p{1.8cm}>{\centering\arraybackslash}p{1.8cm}>{\centering\arraybackslash}p{1.8cm}>{\centering\arraybackslash}p{1.5cm}>{\centering\arraybackslash}p{1.4cm}>{\centering\arraybackslash}p{1.4cm}}
    \hline
    $\mathrm{Kn}_\infty$ & $n_\infty^{\mathrm{He}} \, [\mathrm{m}^{-3}]$ & $n_\infty^{\mathrm{Ar}} \, [\mathrm{m}^{-3}]$ & $n_\infty^{\mathrm{Kr}} \, [\mathrm{m}^{-3}]$ & $U_\infty \, [\mathrm{m/s}]$ & $T_\infty \, [\mathrm{K}]$ & $T_{\mathrm{wall}} \, [\mathrm{K}]$ \\
    \hline
    \hline
    $0.02$& $4.052 \times 10^{19}$ & $6.078 \times 10^{19}$ & $1.013 \times 10^{20}$ & $675.37$ & $200$ & $300$ \\
    \hline
\end{tabular}
\end{table}

\begin{figure*}[t]
	\centering
	\begin{subfigure}{0.485\textwidth}
		\centering
        \includegraphics[width=1.\linewidth]{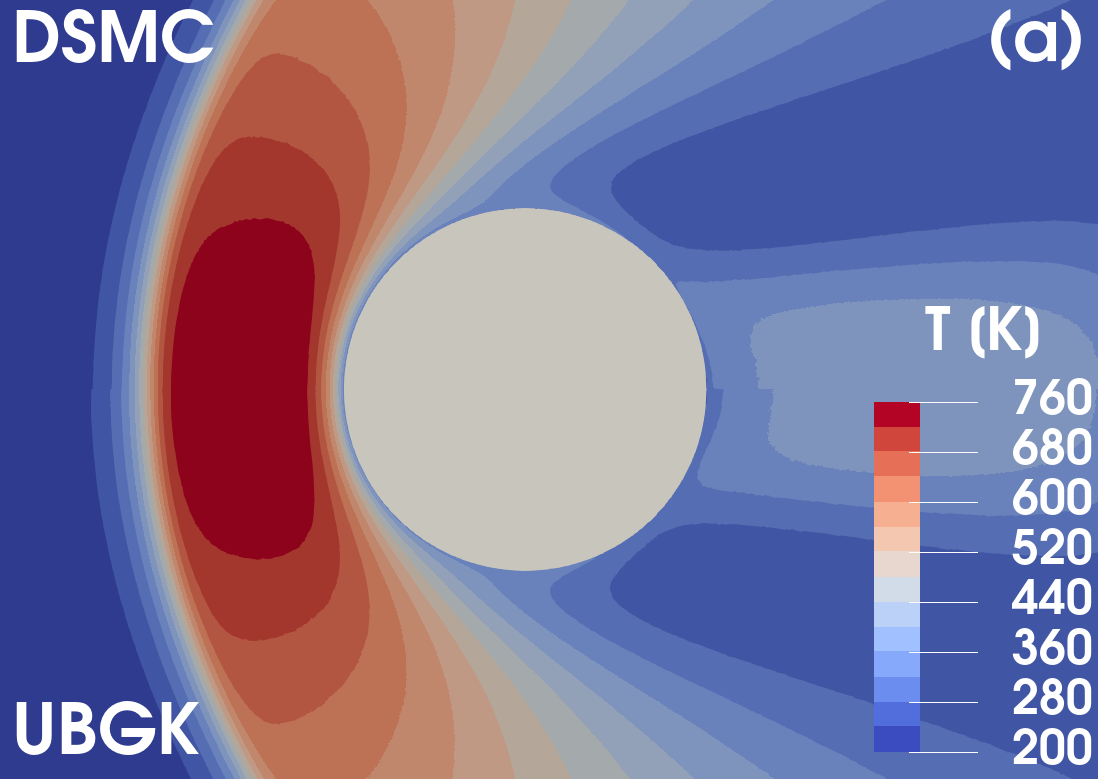}
            \caption{Mixture temperature.}\label{subfig:2_Cyl-Temp-Mix}
	\end{subfigure}
	\begin{subfigure}{0.485\textwidth}
		\centering
        \includegraphics[width=1.\linewidth]{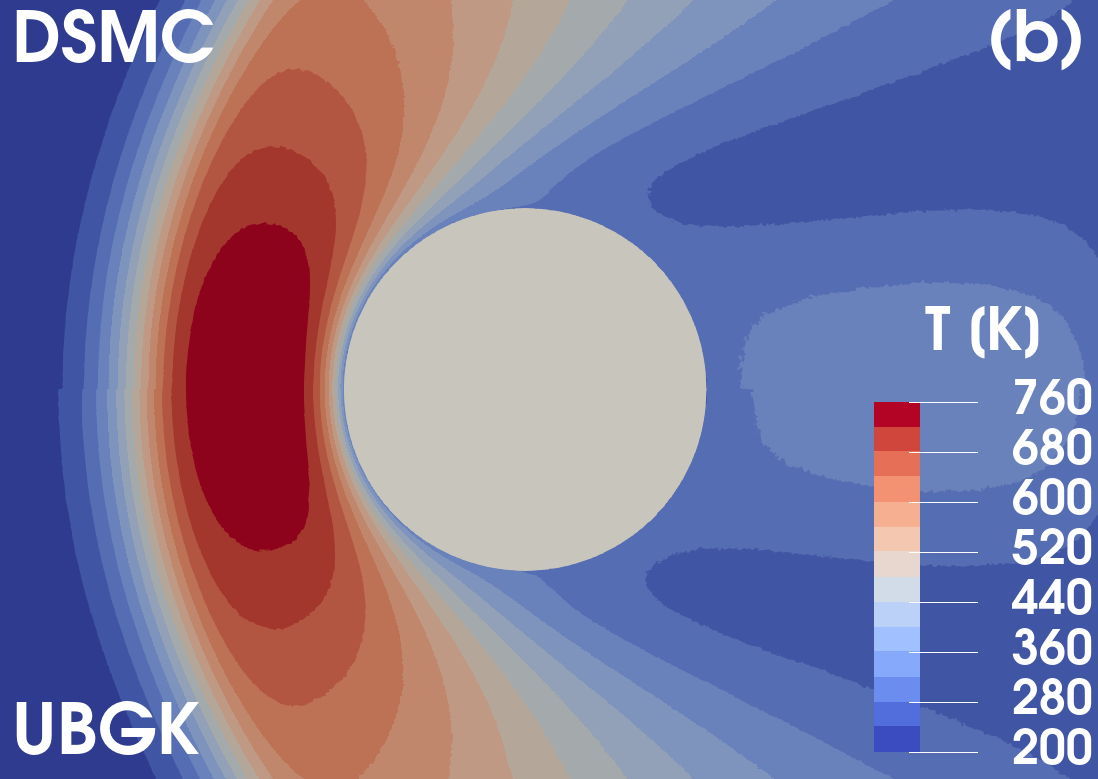}
            \caption{Helium temperature.}\label{subfig:2_Cyl-Temp-He}
	\end{subfigure}
	\centering
	\begin{subfigure}{0.485\textwidth}
		\centering
        \includegraphics[width=1.\linewidth]{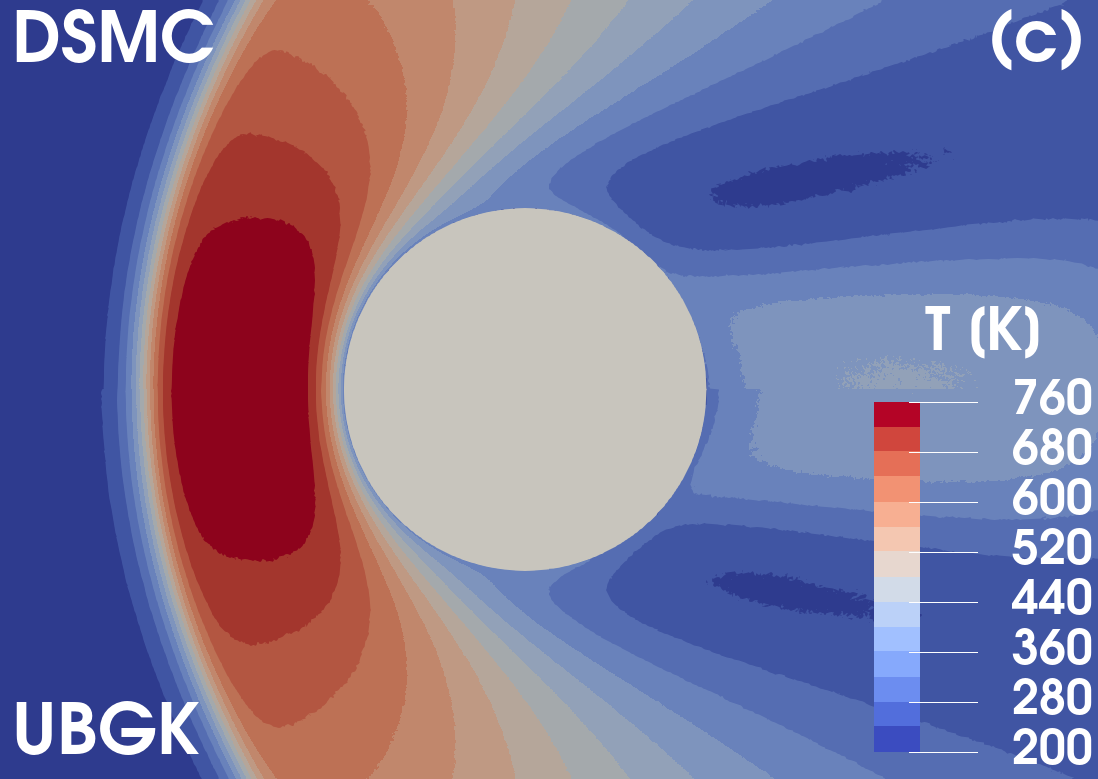}
            \caption{Argon temperature.}\label{subfig:2_Cyl-Temp-Ar}
	\end{subfigure}
	\begin{subfigure}{0.485\textwidth}
		\centering
        \includegraphics[width=1.\linewidth]{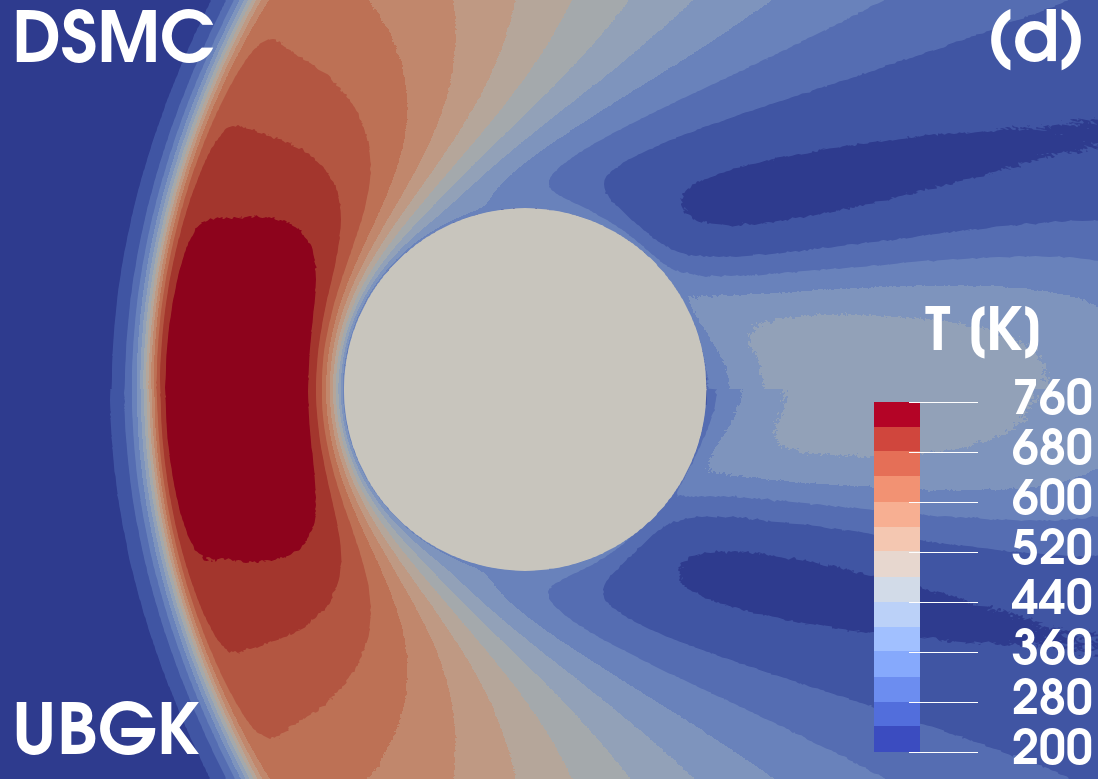}
            \caption{Krypton temperature.}\label{subfig:2_Cyl-Temp-Kr}
	\end{subfigure}
	\caption{Temperature contours for the mixture, Helium, Argon, and Krypton.}\label{fig:2_Cyl-Temp}
\end{figure*}


\begin{figure}[t]
    \centering
	\begin{subfigure}{0.495\textwidth}
		\centering
        \includegraphics[width=1.0\linewidth]{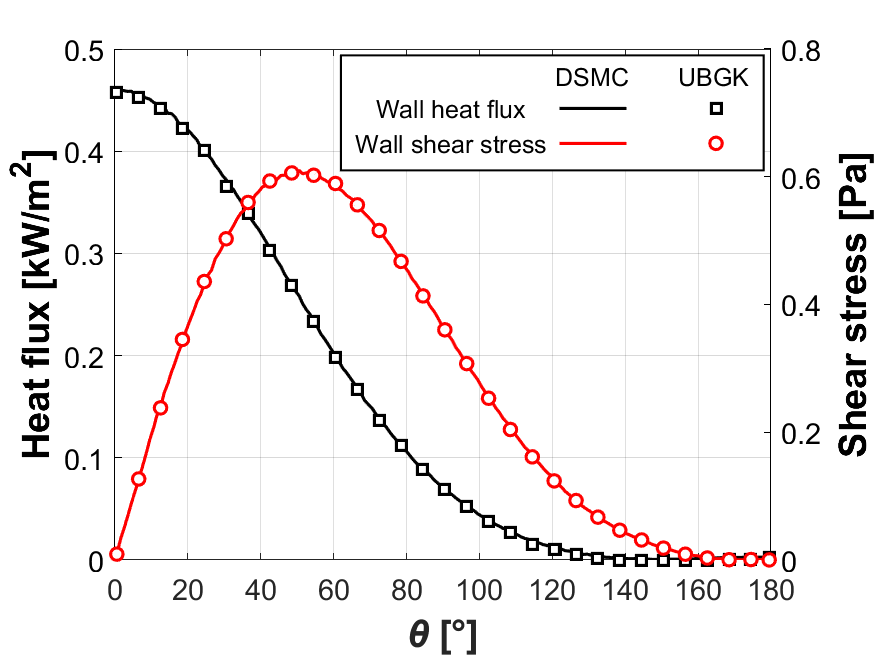}
	\end{subfigure}
    \caption{Wall heat flux and shear stress distribution over a cylinder. The surface angle is measured from the upstream stagnation point.}\label{fig:2_Cyl-HeatShear}
\end{figure}

\begin{table}[t]
\caption{Comparison of peak wall heat flux, total drag, and computation time between DSMC and the UBGK model.}\label{table:2_Cyl-HeatDragPerError}
\centering
\begin{tabular}{>{\raggedright\arraybackslash}p{1.5cm}>{\centering\arraybackslash}p{4cm}>{\centering\arraybackslash}p{3cm}>{\centering\arraybackslash}p{3.2cm}}
    \hline
    Method & Peak wall heat flux $[\mathrm{W/m^2}]$ & Total drag $[\mathrm{N/m}]$ & Computation time $[\mathrm{s}]$ \\
    \hline
    \hline
    DSMC & $460.053$ & $1.8838$ & $40507$ \\
    UBGK & $457.525 \, (0.5495 \, \%)$ & $1.8837 \, (0.0043 \, \%)$ & $46750 \, (15.41 \, \%)$ \\
    \hline
\end{tabular}
\end{table}

Figure \ref{fig:2_Cyl-HeatShear} presents the surface distributions of wall heat flux and shear stress, while Table \ref{table:2_Cyl-HeatDragPerError} provides a quantitative comparison of the peak heat flux and total drag predicted by the UBGK model and DSMC. The computational cost is also summarized in Table \ref{table:2_Cyl-HeatDragPerError}. The computation times were obtained using 192 cores on an AMD EPYC 9654 processor with a clock speed of $2.4 \, \mathrm{GHz}$. Despite the comparable accuracy of surface properties, the computational cost of the UBGK model turned out to be $15.41 \, \%$ higher than that of DSMC. The higher computational cost mainly results from cell-level evaluations of macroscopic quantities and the first-order treatment of the present numerical scheme.

\section{Conclusion}


In this study, a multi-relaxation UBGK formulation is presented for rarefied monatomic gas mixtures. The single-species UBGK framework is extended to account for inter-species interactions, with pair-dependent relaxation properties obtained by enforcing consistency between the UBGK and Boltzmann production terms. The resulting model is implemented in a particle framework and assessed against DSMC solutions. The comparisons show that the model reproduces the temporal evolution of macroscopic properties in homogeneous relaxation and provides accurate predictions of shock-layer structure, surface heat flux, and aerodynamic drag. For future work, the present formulation will be extended to diatomic and polyatomic gas mixtures by incorporating rotational and vibrational energy relaxation.

\begin{credits}

\subsubsection{\ackname} This work was supported by the National Research Foundation of Korea (NRF) grant funded by the Korean government (MSIT) in 2026 (No. RS-2025-02213804, Project Title: In-Space Servicing and Manufacturing Research Center (ISMRC)).

\end{credits}

\bibliographystyle{splncs04}
\bibliography{RGD_UBGK_refs}

\end{document}